\documentclass[preprint,12pt]{elsarticle}

\journal{}

\begin{document}

\begin{frontmatter}

\title{\textbf{A Security Policy Model Transformation and Verification Approach for Software Defined Networking}}

\author[1]{Yunfei Meng}
\author[1]{Zhiqiu Huang}
\author[1]{Guohua Shen}
\author[2]{Changbo Ke}

\address[1]{College of Computer Science and Technology, Nanjing University of Aeronautics and Astronautics, Nanjing 211106, China}
\address[2]{School of Computer Science and Technology, Nanjing University of Posts and Telecommunications, Nanjing 210023, China}

\begin{abstract}
Software defined networking (SDN) has been adopted to enforce the security of large-scale and complex networks because of its programmable, abstract, centralized intelligent control and global and real-time traffic view. However, the current SDN-based security enforcement mechanisms require network managers to fully understand the underlying configurations of network. Facing the increasingly complex and huge SDN networks, we urgently need a novel security policy management mechanism which can be completely transparent to any underlying information. That is it can permit network managers to define upper-level security policies without containing any underlying information of network, and by means of model transformation system, these upper-level security policies can be transformed into their corresponding lower-level policies containing underlying information automatically. Moreover, it should ensure system model updated by the generated lower-level policies can hold all of security properties defined in upper-level policies. Based on these insights, we propose a security policy model transformation and verification approach for SDN in this paper. We first present the formal definition of a security policy model (SPM) which can be used to specify the security policies used in SDN. Then, we propose a model transformation system based on SDN system model and mapping rules, which can enable network managers to convert SPM model into corresponding underlying network configuration policies automatically, i.e., flow table model (FTM). In order to verify SDN system model updated by the generated FTM models can hold the security properties defined in SPM models, we design a security policy verification system based on model checking. Finally, we utilize a comprehensive case to illustrate the feasibility of the proposed approach.
\end{abstract}


\end{frontmatter}

\section{Introduction}

\paragraph{}
Currently software-defined networking (SDN)  \cite{63} has been adopted to enforce the security of large-scale and complex networks because of its programmable, abstract, centralized intelligent control and global and real-time traffic view. For instances, Garay et al \cite{66} proposed a SDN-based network access control mechanism, flownac, which is a centralized EAP (Extensible Authentication Protocol) based on terminal security authentication method for IEEE 802.1x wireless local area network (WLAN). Yakasai et al \cite{67} proposed the flowidentity network access control mechanism. This mechanism integrates EAP security authentication mechanism into SDN controller, and then uses custom firewall policy to control terminal access. Hu et al \cite{68} proposed a dynamic firewall mechanism based on SDN, flowguard. With this mechanism, network administrators can flexibly customize different types of firewall policies and authenticate different types of network devices. Koerner et al \cite{69} proposed a device security authentication mechanism based on MAC address and SDN. This mechanism uses the floodlight controller to map the MAC address of the laptop to the corresponding VLAN address in the network. Because the MAC address is static, it can ensure the security authentication of the mobile workstation. However, current SDN-based security enforcement mechanisms require network managers to fully understand the detailed underlying configurations of network (such as MAC address, IP address, VLAN ID, network type and etc.) and then load these security policies (such as access control policies or firewall policies) containing underlying information of the network into the SDN controller by means of control programs or manual input. However, facing the increasingly complex and huge SDN networks, traditional security policy management is becoming more and more difficult because it is nearly impossible for network managers to fully understand all of underlying configurations. Moreover, with the emergence of multi-controller SDN \cite{70}, network managers need to manage a variety of heterogeneous SDN controllers at the same time. In this case, the same security policy often needs to be developed and deployed for different types of controllers, which inevitably increases the complexity and difficulty of network management. Therefore, in the face of the increasingly complex and huge SDN networks, we urgently need a novel security policy management mechanism which can be completely transparent to any underlying information of network. That is it can permit network managers to define upper-level security policies without containing any underlying information of network, and by means of model transformation system, these upper-level security policies can be transformed into their corresponding lower-level policies containing underlying information automatically. Moreover, it should ensure SDN system model updated by the generated lower-level policies can hold all of security properties defined in upper-level policies.

\paragraph{}
Based on these insights, we propose a security policy model transformation and verification approach for SDN in this paper. The main idea of the approach is to specify the current security policies used in SDN, such as access control policies or firewall policies, as a unified security policy model (SPM). SPM is of an upper-level policy model without containing any underlying information of SDN. Then, we establish SDN system model and use SDN system model to establish the mapping rules between the objects of SPM and the objects of SDN system models, and then use these mapping rules to automatically convert the SPM models into their corresponding lower-level policy models containing underlying information of SDN, i.e., flow table model (FTM). To be practically useful, we must prove that SDN system model updated by the generated FTM can ensure all of security properties defined in SPM after model transformation. Thus, we design a verification algorithm, which validates the SDN system model updated by FTM by means of the security properties defined in SPM, i.e., validation conditions. If all of given validation conditions are proofed to be true after verification, it proves that the updated SDN system model can hold all of security properties defined in SPM, otherwise it cannot satisfy the security requirements. Hence, the main contributions of this paper can be concluded as follows:

\paragraph{}
$\bullet$ We establish the system model of SDN in this paper, which includes terminal model (TM), OpenFlow switch model (SWM), flow table model (FTM), network flow model (NFM) and network topology model (NTM).

\paragraph{}
$\bullet$ Since most of security policies used in SDN can be described as the problem whether the policy subject (user, service or terminal, etc.) can access or use the policy object (resource, service, data or terminal, etc.), thus we specify these security policies as a unified security policy model (SPM) in this paper.

\paragraph{}
$\bullet$ Based on the established SDN system model, we further establish the mapping rules between the objects of SPM and the objects of SDN system models. Then, based on these mapping rules, we propose a model transformation system which can automatically transform SPM model without containing any underlying information of network into its corresponding lower-level policy model containing underlying information, i.e., flow table model (FTM).

\paragraph{}
$\bullet$ In order to prove SDN system model updated by the generated FTM can hold all of security properties defined in SPM after model transformation, we design a security policy verification system based on model checking. The verification algorithm of system validates the SDN system model updated by generated FTM by means of the security properties defined in SPM, i.e., validation conditions. If all of given validation conditions are true, it proves that the updated SDN system model can hold all of security properties defined in SPM, otherwise it cannot satisfy the security requirements.

\paragraph{}
$\bullet$ Finally, we utilize a comprehensive case to illustrate the feasibility of the proposed approach. We first establish SDN system model and SPM of this case, then transform SPM into its corresponding FTM using model transformation algorithm, and verify the generated FTM can hold all of security properties defined in SPM after model transformation using security policy verification algorithm.

\paragraph{}
The remainder of the paper is structured as follows. Section 2 discuss some related works. Section 3 is the main body of this paper, which includes the framework of the proposed approach, SDN system model, security policy model, model transformation system and security policy verification system using model checking. In section 4, we utilize a comprehensive case to illustrate the feasibility of approach. Finally, Section 5 concludes this paper and presents some future directions.

\section{Related Work}

\paragraph{}
In this section, we discuss some research works related with policy model transformation and security policy verification, and compare these works with our approach comprehensively.

\subsection{Policy Model Transformation}

According to the definitions of model driven architecture (MDA), model transformation refers to the process of transforming the platform independent model (PIM) to its corresponding platform specific model (PSM) \cite{71}. In terms of the literatures we have reviewed, researches on policy model transformation may be roughly divided into three categories, they are template-based transformation, RBAC-based transformation, as well as system model $\&$ mapping rules-based transformation \cite{72}. Due to the limitations of the template, template-based transformation \cite{73} has only very limited policy transformation ability. RBAC-based transformation \cite{74} is generally only suitable for RBAC (role-based access control) policy transformation, and does not have ample abilities to describe the underlying system, thus these two kind of methods are not suitable for SDN.

\paragraph{}
System model $\&$ mapping rules-based transformation is an active research issue now. Its main research contents include: (1) \emph{system model}: defines the objects of system and the relationship between system objects; (2) \emph{policy model}: defines the policy object and the relationship between policy objects; (3) \emph{upper and lower level mapping rules}: establish the mapping rules between the upper-level policy objects and lower-level system objects by means of the system model \cite{75}\cite{76}. Based on the system model $\&$ mapping rules, policy model transformation first establish the underlying system model, then establish the mapping rules between the upper-level policy objects and the lower-level system objects based on the system model, and then convert the upper-level policy model to the lower-level policy model according to these mapping rules. In particular, Davy et al \cite{77} proposed a policy model transformation approach based on mapping rules, in which the policy model is defined as a tuple $\langle$ event(E), condition(c), behavior(a), subject(s), object(T) $\rangle$ and ontology is used to establish the mapping relationship among the different system layers. Luck et al \cite{78} proposed a method to transform RBAC model defined in service layer into the policy model used in system layer. In this method, the system model is divided into three layers: roles $\&$ object (RO), subject $\&$ resources (SR) and processes $\&$ hosts (PH), and the mapping rules between the three layers have been constructed. Based on the Luck's research, Porto et al. \cite{79} further decomposes the PH layer into two sub layers, namely DAS (diagram abstract subsystem) layer and PH layer. DAS layer is mainly used to describe the network topology in the original PH layer, while PH layer is used to describe the specific network information in DAS layer. In addition, the authors also proposed a policy verification framework, which can be used to verify the consistency problems in the process of policy transformation. In addition, Lampson et al \cite{80} proposed a network policy model transformation approach for distributed computing environment. Maullo et al \cite{81} proposed a policy transformation system based on the first-order predicate logic, which transforms the high-level policy model into the low-level network configuration policy through network topology and other information. Nanxing et al \cite{82} proposed a SDN-oriented access control policy transformation framework. In this paper, we also propose a security policy model transformation based on system model $\&$ mapping rules. We first establish the SDN system model, then further establish the mapping rules between the objects of security policy model and the objects of system model. Next, we propose a model transformation algorithm based on these mapping rules.

\subsection{Security Policy Verification}

\paragraph{}
To assure the information system running securely, security mechanisms of information system have to be validated to check if they conform to the security policies. Traditional test and simulation based validations can only confirm the system work properly in situations described in the scenarios, but they are difficult to identify hidden flaws that are not likely to occur probabilistically. While formal verification methods have been applied to overcome the traditional methods’ shortcoming mentioned above. Two common formal methods for security policy verification are theorem proving and model checking \cite{Ma}. Theorem proving is unsuitable to validate the complex systems properties due to its lower efficiency. Model checking \cite{Clarke} can be used to validate the system whether it conforms to the property expected by modeling the system’s dynamic behaviors and static properties. Model checking has been widely used in the fields of protocol analysis, hardware design and other situations.

\paragraph{}
Security policy is the core of the protection mechanism in information systems. Many researches on security policy verification have been proposed. For instances, Al-Shaer et al.\cite{Shaer} proposed a static policy inconsistency detection method for firewall policies. Bandara et al.\cite{Bandara} proposed a Event calculus (EC) based framework for security policy verification. They translate policy and system behavior specification to a formal notation based on event calculus and then used reasoning techniques for conflict identification. May et al.\cite{May} verified privacy policies with an asynchronous model checker. Rubio-Loyola et al.\cite{Rubio} proposed a goal-oriented policy refinement method by means of a model checking technique with linear temporal Logic (LTL). Graham et al.\cite{Graham} proposed an incremental method to validate policy based systems through direct conflict detection with extended decision table. Baliosan and Serrat \cite{Baliosian} developed a specific finite automaton based approach for policy conflict detection. In this paper, we also propose a security policy verification system using model checking. We first specify the security properties defined in security policy model as a group of validation conditions, then input validation conditions and SDN system model into the verification algorithm for checking security properties.

\section{Security Policy Model Transformation and Verification Approach}

\paragraph{}
We propose a security policy model transformation and verification approach for software defined networking in this paper. In the following of this section, we first present the framework of approach. Then, we establish SDN system model and present the proposed security policy model (SPM). Next, based on the established SDN system model, we present how to transform SPM model into its corresponding FTM model using mapping rules. After that, we present how to verify the generated FTM can hold all of security properties defined in SPM by means of model checking algorithm.

\subsection{Framework of Approach}

\paragraph{}
We depict the framework of entire approach as Figure 1, which includes security policy model (SPM), model transformation system, security policy verification system and SDN system model. Firstly, we specify the security policies (such as access control policies or firewall policies) used in SDN as a unified SPM. Then, we establish SDN system model. Based on SDN system model, we further establish the mapping rules between the objects of SPM and the objects of SDN system model, and then utilize these mapping rules to automatically convert SPM model into its corresponding underlying network configuration policies, i.e., flow table model (FTM). To be practically useful, we must prove that SDN system updated by the generated FTM can ensure all of security properties defined in SPM after model transformation. Thus, we design a security policy verification system based on model checking technique. Before verification, we first update SDN system model using the generated FTM, and the updated SDN system model is denoted as $RSDN$. Then, we specify the security properties defined in SPM as a group of specific validation conditions $\{\ VC_{i}\ \}$, then we input RSDN and $\{\ VC_{i}\ \}$ into security policy verification system and execute model checking. If all of given validation conditions $\{\ VC_{i}\ \}$ are proofed to be true after verification, it proves that the updated SDN system model can hold all of security properties defined in SPM. Otherwise, it cannot satisfy the security requirements of SPM. In this case, we will redesign our model transformation system until it can meet all of requirements.

\begin{figure}[htbp]
\centering
\scalebox{0.6}{\includegraphics{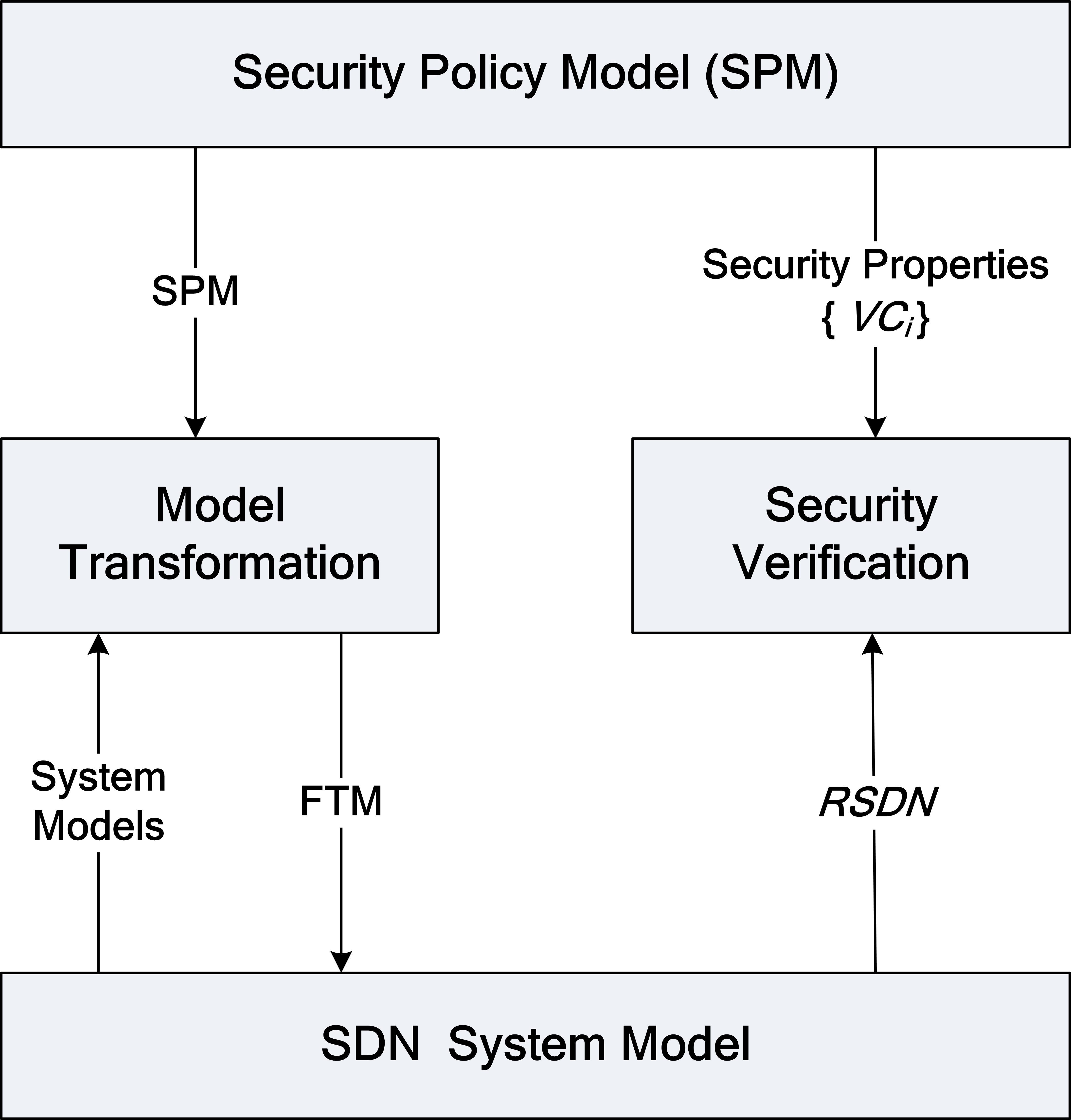}}
\caption{The framework of approach.}
\end{figure}

\section{Conclusion}

\paragraph{}
In order to implement transforming the upper-level security policies without containing any underlying information of SDN into their corresponding lower-level policies containing underlying information automatically, we propose a security policy model transformation and verification approach for SDN in this paper. We first establish SDN system model, which includes terminal model (TM), OpenFlow switch model (SWM), flow table model (FTM), network flow model (NFM) and network topology model (NTM). Since most of security policies used in SDN can be described as the problem whether the policy subject (user, service or terminal, etc.) can access or use the policy object (resource, service, data or terminal, etc.), thus we specify these security policies as a unified SPM in this paper. Based on the established SDN system model and mapping rules, we propose a model transformation system which can automatically transform SPM model into its corresponding lower-level policy model, i.e., FTM. In order to prove SDN system model updated by the generated FTM can ensure all of security properties defined in SPM after model transformation, we design a security policy verification system based on model checking. The verification algorithm validates the SDN system model updated by the generated FTM by means of a group of validation conditions. If all of validation conditions are true, it proves that the updated SDN system model can hold all of security properties defined in SPM, otherwise it cannot satisfy the security requirements. Finally, we utilize a comprehensive case to illustrate the feasibility of the proposed approach.

\paragraph{}
At present, the route selection algorithm $\alpha$ adopted by model transformation system is set as the simplest Dijkstra shortest path search algorithm. But when $\alpha$ is set as some more complex algorithms, such as linear programming, load balancing or quality of services (QoS) aware and etc., the proposed model transformation system needs to be further verified and evaluated. Hence, to further expand our model transformation system will be an interesting direction for our ongoing works.

\section*{Acknowledgments}

\paragraph{}
This paper has been sponsored and supported by National Natural Science Foundation of China (Grant No.61772270), partially supported by National Natural Science Foundation of China (Grant No.61602262).

\section*{References}

\bibliographystyle{elsarticle-num}
\bibliography{02}



%



\end{document}